\documentclass[12pt,a4paper]{article}
\usepackage[utf8]{inputenc}
\usepackage[T1]{fontenc}
\usepackage{authblk}
\usepackage{amsmath}
\usepackage{amsfonts}
\usepackage{amssymb}
\usepackage{graphicx}
\newcommand{\be}{\begin{equation}}
\newcommand{\ee}{\end{equation}}
\newcommand{\bea}{\begin{eqnarray}}
\newcommand{\eea}{\end{eqnarray}}

\title{\vspace{-1.8in} On the Entropy of Strings and Branes}
\author{Ram Brustein, Yoav Zigdon}
\affil{{\normalsize Department of Physics, Ben-Gurion University, Beer-Sheva 84105, Israel} \\
	{\small\tt ramyb@bgu.ac.il\ \ \ \ yoavzig@post.bgu.ac.il}}
\date{}
\begin{document}
	\maketitle
	\begin{abstract}
We show that the entropy of strings that wind around the Euclidean time circle is proportional to the Noether charge associated with translations along the T-dual time direction. We consider an effective target-space field theory which includes a large class of terms in the action with various modes, interactions and $\alpha'$ corrections.
The entropy and the Noether charge are shown to depend only on the values of fields at the boundary of space. The  classical entropy, which is proportional to the inverse of Newton's constant, is then calculated by evaluating the appropriate boundary term for various geometries with and without a horizon. We verify, in our framework, that for higher-curvature pure gravity theories, the Wald entropy of static neutral black hole solutions is equal to the entropy derived from the Gibbons-Hawking boundary term.  We then proceed to discuss horizonless geometries which contain, due to the back-reaction of the strings and branes, a second boundary in addition to the asymptotic boundary. Near this ``punctured'' boundary, the  time-time component of the metric and the derivatives of its logarithm approach zero. Assuming that there are such non-singular solutions, we identify the entropy of the strings and branes in this geometry with the entropy of the solution to all orders in $\alpha'$. If the asymptotic region of an $\alpha'$-corrected neutral black hole is connected through the bulk to a puncture, then the black hole entropy is equal to the entropy of the strings and branes.  Later, we discuss configurations similar to the charged black p-brane solutions of Horowitz and Strominger, with the second boundary, and show that, to leading order in the $\alpha'$ expansion, the classical entropy of the strings and branes is equal exactly to the Bekenstein-Hawking entropy. This result is extended to a configuration that asymptotes to AdS.
%\end{spacing}
	\end{abstract}
\pagebreak
	\tableofcontents
\pagebreak

\section{Introduction}

\linespread{1.5}\selectfont
\setlength{\parindent}{0em}
Black holes possess entropy, the Bekenstein-Hawking entropy \cite{Bekenstein72},\cite{Bekenstein73}, \cite{Hawking}, \cite{GH}, which agrees with certain microscopic counting of bound states of strings and branes that wrap internal cycles \cite{StromingerVafa},\cite{CallanMaldacena},\cite{BMPV},\cite{BLMPSV},\cite{HMS}. The agreement is with the thermodynamic entropy of certain extremal and near-extremal charged black holes, but a calculation of this type is not available for neutral black holes. Given this situation, we appeal to an effective description of strings and branes in order to attempt to explain the entropy of large non-extremal black holes. The idea is that the entropy is the classical thermodynamic entropy of the stringy matter.
	
%A related idea is  that strings and black holes can transition into each other under certain circumstances \cite{Bowick:1985af}, \cite{Susskind93}. The arguments supporting the possibility of such a transition rely on comparing the entropy, temperature and scale of the black hole and the string. These were extended to charged systems in \cite{HP96} (see also \cite{DamourVeneziano}). Motivated by the idea of the black hole/string transition,
Stringy matter can give rise to classical entropy, namely, one that scales like the inverse of Newton's constant, similarly to the Bekenstein-Hawking entropy. For example, Horowitz and Polchinski (HP) \cite{HP97} found a solution featuring a quasi-localized condensate of closed strings that wind around the thermal circle in Euclidean signature. This solution has a classical entropy and its Lorentzian interpretation involves highly-excited, hot and self-gravitating gas of strings.
	
In the HP solution, the scalar field representing the condensate has a Gaussian-like profile and the thermal circle shrinks by a small amount. The equations that this solution satisfies are derived from an effective field theory (EFT) action - the HP action - in which the dominant interaction term between the compact-compact graviton (the radion) and winding modes \cite{AW}\footnote{Note that there, the massless mode whose interaction with the winding modes induces first-order phase transition is the radion.}.

Both small and large (in string units) stringy Euclidean black hole solutions include a closed string winding condensate  \cite{Kutasov:2005rr},\cite{CMW}.  It was pointed out that this condensate has a classical entropy \cite{CM}, which constitutes at least some part of the black hole entropy \cite{CMW}, but it was not clear how much of the entropy is accounted for by the condensate.

As originally suggested by Dabholkar \cite{Atish}, recent papers \cite{WindingEntropy}, \cite{Amit21}, argued that the winding condensate accounts for the entire Bekenstein-Hawking entropy of the black holes,  specifically when taking into account the back reaction it induces \cite{BGIZ}. However, the result seemed too good to be true for several reasons.  It was not clear why all the other winding modes, which become light as the Euclidean time circle pinches off, do not induce significant corrections to the entropy.  Furthermore, the winding condensate varies rapidly over a few units of string length near the tip, therefore one would have expected that $\alpha'$ corrections would give rise to significant corrections to the entropy.\footnote{Additionally, the geometry of the Euclidean black hole is that of cigar with a smooth tip, and the near-tip region contains a factor of $R^2$, making the concept of winding ill defined.}${}^{,}$\footnote{ The application of an EFT of winding modes for a given Euclidean black hole solution (say, Schwarzschild) poses an additional problem.  Far away from the tip, the profile of the winding field is exponentially small and non-perturbative in $\alpha'$. In standard treatments, non-perturbative instantons in field theories are not introduced as fields in an action and their back-reaction is not considered.} One of the goals of this paper is to address these issues.

Recently, several papers were written about winding condensates. We calculated the coefficient of the interaction term between a radion and two winding modes by a string S-matrix computation for type II and the bosonic string in \cite{EFT}. An additional interaction, the quartic interaction between four winding modes, was similarly computed in \cite{EFT} for the bosonic string theory and type II superstring theory. A related calculation appeared in \cite{Dine03}, and in \cite{Troost}, the same interactions were calculated for the Heterotic string. In \cite{Jafferis} it was argued that there is an Einstein-Rosen$=$Einstein-Podolsky-Rosen duality between two-sided black holes and entangled states of folded string pairs on a disjoint union of linear-dilaton Minkowski spacetimes. See \cite{AmitSunny} for an earlier incarnation of the argument. A 3D version of the Fateev-Zamoldchikov-Zamolodchikov (FZZ) duality \cite{KazakovKostovKutasov},  between a WZW model describing AdS$_3$ and a target-space with non-contractible thermal cycle deformed by a winding condensate, received evidence in \cite{Jafferis2}.

In \cite{CM}, expressions for the entropy of winding modes and their profiles for large-D black holes were written, by treating them as small perturbations. In \cite{CMW} a charged version of the HP solution was found and it was argued that in classical type II superstring theory the transition between an HP phase and a black hole phase cannot be smooth. Later, we extended the HP action by including the quartic interaction and NS-NS flux and found solutions in which the thermal circle has a fixed circumference in space. These solutions were interpreted as describing strings in thermal equilibrium slightly above the Hagedorn temperature \cite{ThermalEquilibrium}. In \cite{Yiming}, a variant of the spectral form factor was considered and its increase in time was explained for free string theories by identifying relevant string microstates.

Additionally, the author conjectured the existence of complex HP-like solutions with higher winding and momentum numbers which would explain the expected ramp-up in time of this variant in weakly-coupled string theories.
An HP-like solution with an asymptotically AdS factor of the geometry was found in \cite{Erez}, its instability and a potential transition to a small black hole in AdS were also discussed. The authors of \cite{David} found a family of HP-like solutions that asymptote to $S^1 _{\beta}\times R^6$ where $\beta$ corresponds to the inverse Hagedorn temperature, and also identified worldsheet conformal field theories (CFTs)  that describe them. In \cite{Matsuo}, a solution of the winding modes coupled to Einstein's gravity was found and it was argued that the condensate behaves approximately like a perfect fluid.

Here, we present a general relation between the entropy of winding modes and the Noether charge $Q$, associated with translations in the T-dual of the thermal circle: they are proportional $Q\propto \beta S$, $\beta$ being the inverse temperature. A similar statement holds for the T-dual momentum modes. This can be viewed as a stringy realization of the relationship between the entropy and the Noether charge associated with translations along the thermal circle \cite{Wald},\cite{IyerWald},\cite{Visser} (see also \cite{MyersJac1}, \cite{MyersJac2}), albeit with significant differences. The relationship between the entropy and the Noether charge is shown to hold also for a variety of higher-order correction terms are added to the EFT action. To establish the result,  we utilize a key new ingredient: that the terms where the winding modes appear in the Lagrangian density of the target-space EFT  depend only on powers of the proper length of the thermal circle.  This property allows us to express the entropy as a boundary term and consequently show that the Noether charge depends only on the values of the fields at the boundary of space and as such it is insensitive to details of the solutions in the bulk.

For a neutral black hole solution of a stringy higher-derivative theory of gravity, in case that the string and brane sources are turned off, we verify the equality of the Wald entropy and the entropy derived from a Gibbons-Hawking procedure - which is valid to leading order in the string coupling and to all orders in $\alpha'$ \cite{Tseytlin88},\cite{KazakovTseytlin},\cite{CMW}. This connection was explained in \cite{Wald}, and also mentioned in passing in \cite{CMW}, however the calculation we present did not appear in these references.

Building on our previous entropy calculations we focus on a puncture in the geometry - in the vicinity of which the $\tau-\tau$ metric $G_{\tau\tau}$, and the first derivatives of $\log(G_{\tau \tau})$ approach zero. Figure 1 depicts a part of the geometry of the 2D ``puncture solution'' that was found in \cite{BGIZ}. The puncture is induced by back reaction, when turning on the winding condensate with an asymptotic fallout condition derived from the $SL(2,R)_k/U(1)$ CFT for large $k$. The puncture makes the concept of winding well-defined, unlike the situation in the vicinity of a smooth tip. In a way, the back reaction of the winding modes saves them from a tragic demise. Also, we expect that some worldsheet superconformal field theories describe higher-dimensional versions of the two-dimensional puncture solution.

\begin{figure}[h]
	\begin{center}
		
		\includegraphics[scale=0.5]{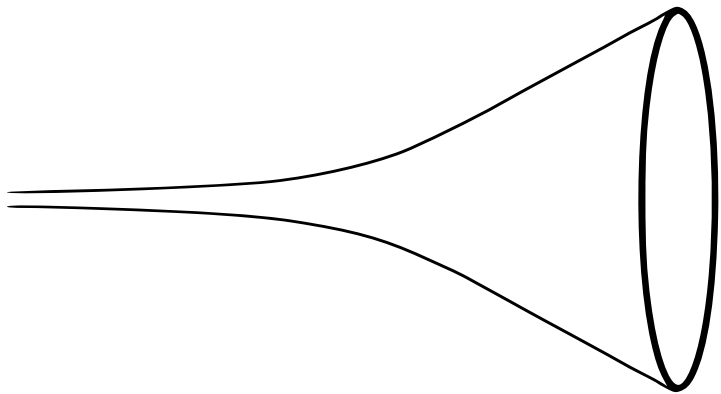}
	\end{center}

\caption{Depicted is the thermal circle as a function of the radial coordinate of the  solution found in \cite{BGIZ}. The approximately linear shrinking of the circumference of the circle on the right occurs on a few string lengths and is shared with the near-tip region of a large Schwarzschild Euclidean black hole. The circumference of the circle and the derivatives of its logarithm approach zero as the left asymptotic boundary is approached, in contrast to the conventional smooth tip. The geometry can be viewed as a Euclidean wormhole. We will consider higher dimensional hypothetical solutions with a 2D factor as above, such that each point in the diagram corresponds to a sphere.}
\end{figure}
We then show that  the entropy of string and brane sources, whose asymptotic geometry is that of a neutral Euclidean black hole and whose inner boundary is punctured, is equal to the Bekenstein-Hawking entropy to leading order in the string coupling and in $\alpha'$. The result holds even when taking into account a large class of $\alpha'$ corrections to the leading order action, provided that the solutions are non-singular. More generally, if the asymptotic geometry is that of an $\alpha'$-corrected black hole, then the entropy derived from the Gibbons-Hawking procedure is identified with that of the string and brane modes.

We extend our argument to geometries that asymptote to those of the black p-branes solutions of Horowitz and Strominger (HS) \cite{HS}, while their inner boundary satisfies the puncture boundary conditions. We show that the leading order entropy of the sources, as computed in the EFT, is equal to the Bekenstein-Hawking entropy of the usual HS solutions. This applies to the neutral, extremal and near-extremal cases.
The same results are obtained for a black p-brane carrying winding charge in a compact circle and also the  $AdS_{D}\times S^D$ asymptotic background with a black hole in the AdS factor and a Ramond-Ramond flux supported on the $S^D$ (for $D=5$). These calculations can also be viewed as  a method to obtain the entropies of the HS solutions  that is simpler than using the Gibbons-Hawking procedure, in particular they do not require a regularization.

The organization of the rest of the paper is as follows. In the next section we show that the Noether charge associated with translations along the T-dual of the thermal circle is proportional to the asymptotic temperature times the entropy of winding condensates. In Section 3 we compute the entropy of strings and branes in various geometries, assuming the regularity of the solutions. In Section 4 we calculate the leading-order entropies of string and brane sources in punctured Horowitz-Strominger geometries and several other geometries. We conclude and discuss the results in Section 5. A short appendix explains a convenient choice of a coordinate system near the horizon of an $\alpha'$-corrected black hole.

\section{Entropy of Strings and the $U(1)$ Noether Charge}

String theory compactified on a circle has a symmetry associated with translations along the circle and along the T-dual circle. On the worldsheet, this is a global symmetry, while in target-space, this is a gauge redundancy. We consider a general target-space EFT which is invariant under these symmetries. Here we show that the  Noether charge $Q$, associated with translations along the T-dual circle,  is proportional to the entropy of the winding-mode condensates and then that the Noether charge $\widetilde{Q}$, associated with translations along the time circle is proportional to the entropy of momentum-mode condensates.  We start with the HP EFT for the winding modes $\pm 1$, and later generalize the discussion by including a large class of terms in the EFT action, corresponding to additional modes and interactions.

Several comments about the charge are in order. First, the idea that entropy is proportional to a Noether charge appeared in  \cite{Wald},\cite{IyerWald}, and its Euclidean version was studied in \cite{Visser}. For static black hole solutions, this entropy was defined as the inverse temperature times a surface integral of a $(D-2)$-form which is the Noether current associated with the $\frac{\partial}{\partial \tau}$ horizon-Killing vector. We show that the connection between the entropy of the winding modes and $Q$ is similar.

The Wald entropy is related to a Noether charge which is evaluated as a surface integral on a cross section of a bifurcating Killing horizon. However, there are some significant differences between the Wald Noether charge the Noether charge $Q$. Our derivation of $Q$ does not rely on the existence of a horizon in the geometry, and in particular, it does not require the existence of a bifurcating Killing horizon. In Wald's construction the vanishing horizon Killing vector is crucial to the derivation.
As we will see in the next section, when considering horizonless configurations, $Q$ is evaluated as a surface integral at infinity. For the standard Euclidean black hole, the charge is evaluated as a sum of surface integrals at infinity and at the horizon.

Also, Wald's derivation is applicable for theories of pure gravity, for which the entropy of winding strings vanishes.  In general, any form of matter does not contribute explicitly to the Wald entropy - it contributes implicitly through its back reaction on the geometry. Our charge is uniquely sourced by winding strings.

The charge $Q$ is positive semi-definite, in contrast to typical gauge charges which could be either positive or negative. In particular, it is not the standard winding charge which can be either positive or negative.
An interpretation of the charge $Q$ is suggested by the comparison of $Q$ to the Lorentzian energy.  The Lorentzian time translation generator can be analytically continued to generate translations along the Euclidean time circle and then can be related to translations along the T-dual $\tilde{\tau}$. The former is the Hamiltonian and the value of the latter is $Q$. Therefore, we interpret the charge $Q$ as the energy of the system.

\subsection{Horowitz-Polchinski Effective Field Theory}

In this subsection we recall the HP effective action which possesses a $U(1)$ symmetry (see Eq.~(\ref{HPAction}))  and rewrite it in order to be able to calculate the associated U(1) current and charge, which we do in the next subsection.

We adopt the following notations:  $\chi$ and $\chi^*$ are fields corresponding to the winding number plus one and minus one modes, respectively. The metric $G_{\mu \nu}$ determines the geometry of the $d$ spatial dimensions, the Euclidean time-time component of the metric is denoted by $G_{\tau \tau}=e^{2\sigma}$, the $d$-dimensional dilaton $\Phi_d $ is related to the $D=d+1$ dimensional dilaton by $\Phi_d = \Phi_D - \frac{\sigma}{2}$, $\beta$ denotes the asymptotic circumference of the thermal circle, $\beta_H$ is the inverse Hagedorn temperature, which depends on the particular string theory, and finally $\frac{1}{\kappa_0^2}$ is a standard tree-level normalization in the string frame \cite{PolchinskiVolI}.
Our starting point is the HP action,
	\begin{equation}
\label{HPAction}
		I_{HP} = \beta \int d^d x \sqrt{G_d} e^{-2\Phi_d} \left( G^{\mu \nu}\partial_{\mu} \chi \partial_{\nu} \chi^* + \frac{\beta ^2 e^{2\sigma} - \beta_H ^2}{(2\pi \alpha')^2}\chi \chi^*\right)+I_{DG}.
	\end{equation}
Here, $I_{DG}$ is the standard dilaton-gravity action,
\begin{equation}
	I_{DG} = -\frac{\beta}{2\kappa_0 ^2}\int d^d x \sqrt{G_d} e^{-2\Phi_d} \left( R_d -G^{\mu \nu}\partial_{\mu} \sigma \partial_{\nu} \sigma+4G^{\mu \nu} \partial_{\mu} \Phi_d \partial_{\nu} \Phi_d\right).
\end{equation}
Reliable solutions of the HP action have the fields $\chi,\chi^*$ light, small, with derivatives that are small in string units, small string coupling and weak curvature in string units.

Recall that the vertex operator of the winding mode with winding number $w$ for a background that contains a fixed $S^1$ of radius $R$ is of the following form,
	\begin{equation}
		V_w \propto e^{i k_L \cdot X_L (z) + ik_R \cdot X_R (\bar{z})} ~,~ k_L=-k_R = w\frac{R}{\alpha'} ~,~ 		R = \frac{\beta }{2\pi}.
	\end{equation}
	In particular,
	\begin{equation}
\label{VertexOperator}
		V_1 \propto e^{i \frac{\beta }{2\pi \alpha'} \tilde{\tau}} ~,~ \tilde{\tau}=\tau_L-\tau_R.
	\end{equation}
		Thus, the corresponding target-space fields take the form:
	\begin{equation}
\label{VertexOperator2}
		\chi = \chi (\vec{r}) e^{i\frac{\beta }{2\pi \alpha'}\tilde{\tau}} ~,~ \chi ^* = \chi^*  (\vec{r})e^{-i\frac{\beta }{2\pi \alpha'}\tilde{\tau}},
	\end{equation}
where $\vec{r}$ is a spatial vector. To ensure periodicity of the fields around the dual of the $T$-dual of the thermal circle, the range of $\tilde{\tau}$ is $0\le \tilde{\tau} \le \frac{(2\pi)^2 \alpha'}{\beta}$. For the  target-space fields corresponding  to momentum modes, the relevant phase is $e^{\frac{2\pi i n \tau}{\beta}}$, $n$ being the momentum number.

%\subsection{Covariant $\tilde{\tau}$ Kinetic Term}

Next, we introduce an integration over  $\tilde{\tau}$, which allows us to recast the ``HP term'', proportional to $\beta^2 e^{2\sigma} \chi \chi^*$, as a covariant $\tilde{\tau}$-kinetic term of $\chi$, as explained below. Later, we will derive from this term a component of a $U(1)$ current, associated with $\tilde{\tau}$ translations and the corresponding $U(1)$ charge.

First, the following average is equal to one:
	\begin{equation}
		\frac{\beta}{(2\pi)^2 \alpha'} \int_0 ^{\frac{(2\pi)^2 \alpha'}{\beta}} d\tilde{\tau} ~=1.
	\end{equation}
Then, the metrics in the $\tau$-frame and the T-dual $\tilde{\tau}$-frame are related by one of the Buscher rules \cite{Buscher}:
	\begin{equation}
\label{busch}
		G_{\tau \tau} = G^{\tilde{\tau}\tilde{\tau}}.
	\end{equation}
It follows that
	\begin{equation}
		\beta\sqrt{G_{\tau \tau}}\frac{\beta }{(2\pi)^2 \alpha'} \int_0 ^{\frac{(2\pi)^2 \alpha'}{\beta}} d\tilde{\tau} ~  \sqrt{G_{\tilde{\tau}\tilde{\tau}}}G^{\tilde{\tau} \tilde{\tau}} \partial_{\tilde{\tau}} \chi \partial_{\tilde{\tau}} \chi ^* =\beta\frac{\beta^2 e^{2\sigma}}{(2\pi \alpha')^2}\chi \chi^*.
	\end{equation}
This gives rise to the HP term:
	\begin{eqnarray}
\label{HPterm}
		&I_1=\beta  \int d^d x \sqrt{G_{\tau \tau}}   \sqrt{G_d}e^{-2\Phi_D} \frac{\beta^2 e^{2\sigma}}{(2\pi \alpha')^2} \chi \chi^*=\nonumber \\
		&=\beta \int d^d x \sqrt{G_{\tau \tau}}   ~\frac{\beta }{(2\pi)^2 \alpha'}\int_0 ^{\frac{(2\pi)^2 \alpha'}{\beta}}   d\tilde{\tau} \sqrt{G_{\tilde{\tau}\tilde{\tau}}} ~  \sqrt{G_d} e^{-2\Phi_d} G^{\tilde{\tau} \tilde{\tau}} \partial_{\tilde{\tau}} \chi \partial_{\tilde{\tau}} \chi ^*.
	\end{eqnarray}
	One can use the equality $\sqrt{G_{\tau \tau}} \sqrt{G_{\tilde{\tau}\tilde{\tau}}} =1$ to further simplify the expression.

The standard spatial kinetic term of $\chi$ is given by:
	\begin{equation}
		I_2=\beta  \int d^d x ~\frac{\beta }{(2\pi)^2 \alpha'}  \int_0 ^{\frac{(2\pi)^2 \alpha'}{\beta}} d\tilde{\tau} ~ \sqrt{G_d} e^{-2\Phi_d} G^{\mu \nu } \partial_{\mu} \chi \partial_{\nu} \chi ^*.
	\end{equation}
	The last term in the action is related to $\beta_H^2$ associated with the mass-squared of $\chi$:
	\begin{equation}
		I_3=-\beta \int d^d x \frac{\beta}{(2\pi)^2 \alpha'} \int_0 ^{\frac{(2\pi)^2 \alpha'}{\beta}} d\tilde{\tau} \sqrt{G_d} e^{-2\Phi_d}\left(	\frac{\beta_H }{2\pi\alpha'}\right)^2\chi \chi^*.
	\end{equation}

The complete HP action is given by
	\begin{equation}
		I_{HP} = I_1+ I_2 + I_3+I_{DG}.
	\end{equation}
A similar calculation appeared in \cite{Mertens}.

\subsection{$U(1)$ Current, Charge and Entropy of Winding Modes at Leading Order}

The goal of this section is to derive a charge associated with the $U(1)$ translation transformations of $\chi$ and $\chi^*$  and to connect it with the entropy of $\chi,\chi^*$. This is done when considering the leading order action in $\alpha'$ for weakly-curved solutions.

In the previous subsection we introduced the $\tilde{\tau}$-part of the kinetic term,
	\begin{eqnarray}
	\label{}
	I_1&=&\beta  \int d^d x~ \sqrt{G_{\tau \tau}}   \sqrt{G_d}e^{-2\Phi_D} \frac{\beta^2 e^{2\sigma}}{(2\pi \alpha')^2} \chi \chi^*=\nonumber \cr
	&=&\beta \int d^d x \sqrt{G_{\tau \tau}}   ~\frac{\beta }{(2\pi)^2 \alpha'}\int_0 ^{\frac{(2\pi)^2 \alpha'}{\beta}}   d\tilde{\tau} \sqrt{G_{\tilde{\tau}\tilde{\tau}}} ~  \sqrt{G_d} e^{-2\Phi_d} G^{\tilde{\tau} \tilde{\tau}} \partial_{\tilde{\tau}} \chi \partial_{\tilde{\tau}} \chi ^* \cr 
&=&\beta \int d^d x~\sqrt{G_d}~ e^{-2\Phi_d}~ G^{\tilde{\tau} \tilde{\tau}} \partial_{\tilde{\tau}} \chi \partial_{\tilde{\tau}} \chi ^*.
\end{eqnarray}
Viewing the $U(1)$ as a global symmetry under which $\chi\to \chi e^{i \frac{\beta}{2\pi\alpha'} \delta \tilde{\tau}}$ and applying the Noether theorem to $I_{HP}$ (where only $I_1$ and $I_2$ are relevant for the computation), a manifest $\tilde{\tau}$-component to the current arises, $J_{\tilde{\tau}} \sim \chi^* \partial_{\tilde{\tau}} \chi -\chi \partial_{\tilde{\tau}} \chi^*$. Both the $\tilde{\tau}$ and the spatial components of the current are therefore given by a single expression,
	\begin{equation}\label{preCurrnent}
		J_{\mu} =  i~ C_D~\frac{\beta}{2\pi \alpha'}   e^{-2\Phi_d}  \left(\chi \partial_{\mu}\chi^* - \chi^* \partial_{\mu} \chi\right).
	\end{equation}
The current is determined up to a $D$-dependent numerical constant that we denote by $C_D$. We will fix its value in the next section to be $\frac{D-2}{D-3}$ for $D>3$ by identifying the charge with the energy.

The current is then given by
	\begin{equation}
\label{Current}
		J_{\mu} =   C_D~\frac{\beta  }{\pi\alpha'} e^{-2\Phi_d} \chi \chi^*\partial_{\mu} \arg(\chi) .
	\end{equation}
The spatial components of the current vanish for solutions in which the NS-NS field $H_3=dB_2$, vanishes. In general they give rise to quantized $H_3$ charges. The $\tilde{\tau}$-``timelike component'' of the current does not vanish,
\begin{equation}
\label{ExpressionForCurrent}
		J_{\tilde{\tau}} =  C_D~ \frac{2  \beta^2 }{(2\pi\alpha')^2} e^{-2\Phi_d} \chi \chi^*.
\end{equation}
The charge associated with this charge density is given by
	\begin{equation}
\label{ChargeEntropy}
		Q = \int d^d x~\sqrt{G_d}~ J^{\tilde{\tau}} = C_D \int d^d x\sqrt{G_d} e^{-2\Phi_d}\frac{2\beta ^2 e^{2\sigma}}{(2\pi \alpha')^2} \chi \chi^*,
\end{equation}
where we used $G^{\tilde{\tau}\tilde{\tau}}=e^{2\sigma}$ as in Eq.~(\ref{busch}). Since the constant $C_D$ is positive, Eq. (\ref{ChargeEntropy}) implies that $Q\geq 0$, in contrast to usual gauge charges which can be positive or negative. Indeed, swapping $\chi$ and $\chi^*$ does not change the $\tilde{\tau}$-component  of Eq. (\ref{ExpressionForCurrent}). The conservation of charge $ \partial_{\tilde{\tau}} Q=0$, is automatic, because  $Q$ is independent of $\tilde{\tau}$.

We now relate the Noether charge $Q$ to the entropy of the winding modes. The relation between thermodynamic entropy and the free energy in Eq. (\ref{HPAction}),
\begin{equation}
\label{SIrelation}
S= \left(\beta\frac{\partial}{\partial\beta} -1\right) I_{HP},
\end{equation}
implies that
\begin{equation}
\label{swind}
	S =  \int d^d x\sqrt{G_d} e^{-2\Phi_d}\frac{2\beta ^3 e^{2\sigma}}{(2\pi \alpha')^2} \chi \chi^*+S_{bdy}.
\end{equation}
The first term on right hand side arises from explicit dependence on $\beta$, which appears in the HP term Eq. (\ref{HPterm}). The equations of motion and the chain rule imply that a contribution from the implicit dependence of the fields on $\beta$ vanishes. The entropy of a generic solution can receive additional contributions from boundary terms, which we denote by $S_{bdy}$~\footnote{We thank Kostas Skenderis for raising this point.}.
An example in which the spatial integral in the R.H.S. of Eq.~(\ref{swind}) vanishes and  $S_{bdy}$ plays a role, is the standard Schwarzschild solution in the absence of winding modes. In this case, the action contains a Gibbons-Hawking-York (GHY) term and the associated boundary term is equal to the Bekenstein-Hawking entropy.
If the boundary terms vanish, which for instance occurs for the HP solution \cite{HP97} and the puncture solution of \cite{BGIZ}, then only the spatial integral contributes. In this case it follows from Eqs.~(\ref{ChargeEntropy}) and (\ref{swind}) that,
\begin{equation}
S=\tfrac{1}{C_D}\beta Q.
\end{equation}
As previously noted, this is analogous to the relation between the Wald entropy of a black hole solution with a Killing symmetry in a diffeomorphism-invariant theory of gravity and  the associated Noether charge:
\begin{equation}
	S_{\text{Wald}} = \beta Q_{\text{Killing}}.
\end{equation}
In spite of this similarity between the charge $Q$ and Wald's Noether charge, there are significant differences which were listed in the beginning of the section.

\subsection{Current, Charge and Entropy of Winding Modes Beyond Leading Order}

In this subsection we wish to calculate the $U(1)$ Noether charge $Q$, for a general target-space action
\begin{equation}
I =\beta \int d^d x \sqrt{G_d} e^{-2\Phi_d} L(\chi_w,\sigma,G_{\mu\nu}),	
\end{equation}
where the fields $\chi_w$ correspond to states with generic winding and momentum numbers. We then calculate the entropy of all the winding modes and relate it to $Q$.

We wish to express a general $U(1)$-invariant term in the Lagrangian density of the winding modes. To this end it is useful to consider first the T-dual momentum modes $T_n,T^* _n$ and then use $T$-duality to deduce the form of a generic winding-mode term.
A general covariant term in the EFT Lagrangian density of the fields $T_n$ has the following form,
\begin{equation}\label{Term123}
	\Delta \widetilde{L} = \left(G^{\tau\tau }\partial_{\tau} T_n \partial_{\tau} T_n ^*\right) ^{n_1}\left(G^{\mu \nu}\partial_{\mu} T_n \partial_{\nu} T _n ^*\right)^{n_2} (T_n T_n ^*)^{n_3} f(\phi_i,\partial_{\mu} \phi_i),
\end{equation}
where $\phi_i$ are additional fields which do not vibrate or wind around the time circle. When writing equations representing terms in the action, such as Eq. (\ref{Term123}), the symbol $\partial_{\mu}$ denotes a covariant derivative. Replacing or adding $\left(G^{\tau\tau }\partial_{\tau} T_n \partial_{\tau} T_n^*\right) ^{n_1}\left(G^{\mu \nu}\partial_{\mu} T_n \partial_{\nu} T_n ^*\right)^{n_2}$ by $\left(G^{\tau\tau }\partial_{\tau} T_n \partial_{\tau} T_n ^*+G^{\mu \nu}\partial_{\mu} T_n \partial_{\nu} T_n ^*\right)^{n}$ would not modify the final result of this subsection.
Including the Euclidean time dependence, as before, $T_n=T_n (\vec{r}) e^{\frac{2\pi i n\tau}{\beta}}~,~T_n^* = T_n ^*(\vec{r}) e^{\frac{-2\pi i n\tau}{\beta}} $,
\begin{equation}
	\Delta \widetilde{L} = \left(G^{\tau\tau }n^2\frac{4\pi^2}{\beta^2} T_n T_n ^*\right) ^{n_1} \left(G^{\mu \nu}\partial_{\mu} T_n \partial_{\nu} T_n ^*\right)^{n_2} (T_n T_n ^*)^{n_3} f(\phi_i,\partial_{\mu} \phi_i).
\label{tildeL}
\end{equation}

We need to apply the T-duality transformations to Eq.~(\ref{tildeL})
\begin{equation}
\label{Relations}
	G^{\tau \tau} \to G_{\tau \tau} ~,~ \beta \to \frac{(2\pi)^2 \alpha'}{\beta} ~,~ T_n \to \chi_w.
\end{equation}
The first relation is a known Buscher rule \cite{Buscher}, the second relation can be understood by writing $R = \frac{\beta}{2\pi}$ and then applying the standard $R \to \frac{\alpha'}{R}$ T-duality transformation. The last relation, with $n=w$, is the conventional interchange between momentum modes and winding modes under T-duality.
The resulting term in the winding-mode Lagrangian $\Delta L$, is given by
\begin{equation}
\label{GeneralDeltaL}
	\Delta L = \left(G_{\tau\tau }\frac{w^2\beta^2}{(2\pi \alpha')^2} \chi_w \chi_w ^*\right) ^{n_1}\left(G^{\mu \nu}\partial_{\mu} \chi_w \partial_{\nu} \chi_w ^*\right)^{n_2} (\chi_w \chi_w^*)^{n_3} f(\phi_i,\partial_{\mu} \phi_i).
\end{equation}

Eq. (\ref{GeneralDeltaL}) is an important equation which we will consider again in the next section. The inclusion of such terms allows one to consider potential solutions in which multiple winding modes become light in some region of the manifold, plus the fields and their derivatives need not be small.

The correction term $\Delta L$, contains the following factor,
	\begin{equation}\label{Term}
		\Delta L \propto \left( \partial_{\mu} |\chi_w|\partial^{\mu} |\chi_w| + |\chi_w|^2 \partial_{\mu} \text{arg} (\chi_w) \partial^{\mu} \text{arg}(\chi_w)\right) ^{n}.
	\end{equation}
Treating the $U(1)$ as if it were a global symmetry, we obtain the contribution to the Noether current from  $\pm w$-winding  modes,
\begin{eqnarray}
	J_w ^{\mu} = i~C_D~\frac{\beta}{2\pi \alpha'} e^{-2\Phi_d} 	\left(w\chi_w \frac{\delta L}{\delta (\partial_{\mu} \chi_w)}-w\chi^* _w \frac{\delta L}{\delta (\partial_{\mu} \chi^* _w)}\right).
\end{eqnarray}
When taking the variations, quantities like $|\chi_w|~,~ \text{arg}(\chi_w)$ are held fixed.
We would like to calculate the $U(1)$ charge.  Observing that
\begin{equation}
	\partial_{\mu} \chi_w = \left(\partial_{\mu} |\chi_w|+i|\chi_w|\partial_{\mu}\text{arg}(\chi_w) \right)e^{i\text{arg}(\chi_w)},
\end{equation}
the chain rule implies that
\begin{equation}
	\chi_w \frac{\delta L}{\delta \left(\partial_{\mu} \chi_w\right)}=\frac{1}{2}|\chi_w| \frac{\delta L}{\delta \left(\partial_{\mu} |\chi_w|\right)}-\frac{1}{2}i\frac{\delta L}{\delta \left(\partial_{\mu}\text{arg}(\chi_w)\right)}.
\end{equation}
The factors of $\frac{1}{2}$ can be obtained by varying the term in Eq. (\ref{Term}).
It follows that
\begin{equation}
\label{Jcurrent}
	(J_{w})^{\mu} = C_D~\frac{\beta}{2\pi \alpha'} e^{-2\Phi_d}   w \frac{\delta L}{\delta \left(\partial_{\mu}\text{arg}(\chi_w)\right)}.
\end{equation}
This is consistent with the leading-order expression in Eq. (\ref{Current}).
 One has
\begin{equation}\label{Phase}
	\partial_{\tilde{\tau}} \text{arg}(\chi_w) = \frac{w\beta }{2\pi \alpha'},
\end{equation}
and by substituting Eq.~(\ref{Phase}) into Eq.~(\ref{Jcurrent}),
we obtain an expression for the charge by integrating over space with the appropriate measure,
\begin{equation}
\label{qu1}
	Q = \int d^d x \sqrt{G_d} \sum_w J^{\tilde{\tau}} _w =  C_D\int d^d x  \sqrt{G_d} e^{-2\Phi_d} \beta \frac{\delta L}{\delta \beta}.
\end{equation}
In this equation, the non-vanishing derivative with respect to $\beta$ comes purely from the explicit dependence on $\beta$, which appears in interactions involving winding modes.
Next, the entropy of winding modes that wrap around the thermal circle comes from the same terms and a boundary term,
\begin{equation}
\label{sbeta}
	S=\int d^d x \sqrt{G_d} e^{-2\Phi_d} \beta^2  \frac{\delta L}{\delta \beta}+\beta^2\int d^{d-1}x \sqrt{G_d} e^{-2\Phi_d} n_{\mu}\sum_w \frac{\delta L}{\delta \left( \partial_{\mu}\chi_w\right)} \frac{\partial \chi_w}{\partial \beta}.
\end{equation}
We consider solutions for which such a term vanishes, this occurs generally for asymptotically flat spaces where the normal derivatives of $\chi_w$ vanish at the boundary.
The combination of Eqs.~(\ref{qu1}), (\ref{sbeta}) implies that
\begin{equation}\label{QbetaS}
S= \tfrac{1}{C_D} \beta Q.
\end{equation}
So far we discussed explicitly a target-space EFT that results from tree-level string theory. However, our derivation does not seem to be sensitive to the dilaton prefactor in the action and so we expect it to be valid also order by order in the string coupling. In this situation the argument applies to $L=L(\chi_w,\sigma,G_{\mu \nu},\Phi_d)$. In addition, performing an S-duality on the system of fundamental strings we started with,  results in a system of D1 branes winding about the thermal circle and we can see that Eq. (\ref{QbetaS}) applies also to this case.

\subsection{Current, Charge and Entropy of Momentum Modes}

We can apply a similar approach to the ${\tau}$-translation Noether charge.
Momentum modes depend on Euclidean time as
\begin{equation}
	T_n = T_n(\vec{r}) e^{\frac{2\pi n}{\beta} i \tau} ~,~ 	T_n^* = T_n^*(\vec{r}) e^{-\frac{2\pi n}{\beta} i \tau},
\end{equation}
The action is just the $T$-dual of the winding modes action, with the replacements
\begin{equation}
	G_{\tau \tau} \to G^{\tau \tau} ~,~ \beta \to \tilde{\beta}= \frac{(2\pi)^2 \alpha'}{\beta}.
\end{equation}
Also,	$ \beta \partial_{\beta} \to -\beta \partial_{\beta}.$

In order to compare to the previous subsection, it is convenient to T-dualize the time circle which originally had the asymptotic circumference $\beta$ and winding modes that depended on $e^{i \frac{\beta w}{2\pi \alpha'}\tilde{\tau}}$, which give rise to  an asymptotic circumference $\frac{(2\pi)^2 \alpha'}{\beta}$ and momentum modes that depend on $e^{i \frac{2\pi n}{\beta}\tau}$. Our arguments below do not rely on T-duality, we only use it to compare with the calculations of the previous subsection.
A general expression for the action is
\begin{equation}
	I = \int_0 ^{\frac{(2\pi)^2 \alpha'}{\beta}} d\tau \int d^d x \sqrt{G_d} e^{-2\Phi_d} L(T_n,G_{\mu \nu},\sigma).
\end{equation}
Thus,  the entropy takes the form
\begin{equation}\label{MomentumEntropy}
	\tilde{S} = (-\beta \partial_\beta -1) I = -\beta \int d^D x \sqrt{G_d} e^{-2\Phi_d} \frac{\delta L}{\delta \beta}.
\end{equation}
The derivative of $\beta$ from the upper limit of the $\tau$ integral cancels with the $-I$ term. As previously, we have in mind cases where additional boundary terms vanish.

Next, let us treat the $U(1)$ gauge redundancy as if it were a global symmetry in order to compute the associated charge. In this case, the current is given by
\begin{equation}
	J^{\mu} = i~ C_D~ \frac{2\pi }{\beta} e^{-2\Phi_d}\sum_n n \left( \frac{\delta L}{\delta \left( \partial_{\mu} T_n\right)} T_n-\frac{\delta L}{\delta \left( \partial_{\mu} T_n ^*\right)} T_n ^*  \right).
\end{equation}
Repeating a step that was previously performed, $J^\mu$ can be rewritten as
\begin{equation}
	J^{\mu} = C_D~\frac{2\pi}{\beta} e^{-2\Phi_d}\sum_n n \frac{\delta L}{\delta \left( \partial_{\mu}\text{arg}(T_n)\right)}.
\end{equation}
As a check, this equation can be reproduced by T-dualizing Eq. (\ref{Jcurrent}) in the previous subsection.
Since $\partial_{\tau} \text{arg}(T_n) = \frac{2\pi n}{\beta}$, the time component of this is
\begin{equation}\label{Jmomentum}
	J^{\tau} = C_D~\frac{1}{\beta} e^{-2\Phi_d}  \frac{\delta L}{\delta \left( \frac{1}{\beta}\right)}=-C_D~ \beta  e^{-2\Phi_d}  \frac{\delta L}{\delta \beta}.
\end{equation}
Consequently, the charge associated with some constant $\tau$ slice is given by
\begin{equation}
\label{ChargeT}
	\widetilde{Q} =-C_D~\beta  \int d^d x \sqrt{G_d}e^{-2\Phi_d}   \frac{\delta L}{\delta \beta}
\end{equation}
and similarly to the previous subsection we obtain
\begin{equation}
\label{ChargeT1}
	\widetilde{S}=\tfrac{1}{C_D}~\tilde{\beta} \widetilde{Q}.
\end{equation}
For static black holes, the Wald entropy is defined as the inverse temperature times Noether current of the symmetry generated by the $\frac{\partial}{\partial \tau}$ Killing vector, integrated over the Killing horizon \cite{Wald},\cite{IyerWald},\cite{Visser}. Equation~(\ref{ChargeT1}) expresses a similar relation between the entropy of momentum modes and the Noether charge. We chose the inverse temperature to be $\tilde{\beta}$ rather than $\beta$ and the two are related by T-duality. The charge $\widetilde{Q}$ is obtained by applying a T-duality transformation to $Q$. We view the results of this section as an explicit stringy realization of the idea of the entropy as a Noether charge.

 \section{Entropy of Strings and Branes - Neutral Cases}

We would like to point out that similarly to fundamental strings, branes can wind around $S^1 _{\beta}$ and one can think of an effective action for them that has terms with non-trivial $\beta$-dependence. For example, one can apply an S-duality on the HP action, transforming  $\chi$ into a winding mode of a D1-brane, with mass squared $m^2\propto \beta^2$. Solutions from such an action are reliable when the string coupling is large and $\alpha'$ corrections are suppressed. For a D$-(p+1)$ brane in $S^1 _{\beta} \times T^{p}$ which winds $w$ times around the thermal circle the mass squared is given by \cite{PolchinskiTASI}
\begin{equation}\label{MassSquaredBrane}
	m^2 _{\text{brane}} = \frac{\pi}{236 \kappa ^2} (4\pi^2 \alpha') ^{10-p} \text{Vol}\left(T^{p}\right) ^2 \beta^2 w^2 ~,~ \kappa^2 = 8\pi G_N.
\end{equation}
This implies that the observation made in \cite{CM} that non-trivial $\beta$ dependence for fundamental strings gives rise to classical entropy is valid also for branes.

In this section we would like to compute the classical entropy of string and brane sources, for a general solution that asymptotes to $S^1 _{\beta} \times R^{d}$ and does not carry charges associated with fluxes. We focus on three types of possible geometries:

\begin{itemize}
	\item Generalized Horowitz-Polchinski geometries, with asymptotic circumference $\beta$ greater than $\beta_H$, in which the thermal circle does not shrink to zero and such that the minimal size of the $S^1$ is not parametrically smaller than $\beta$. Also, an $S^{D-2}$ factor of the manifold shrinks to zero at the origin.

	\item Euclidean Black Holes, in which the thermal circle shrinks to zero at a smooth tip and the radius of the $(D-2)-$ sphere  at the tip is large in string units. While we write general expressions of contributions to the entropy, our main goal in this case is to check an equality between the entropy derived from the Gibbons-Hawking procedure and the Wald entropy when string and brane sources are absent.

	\item ``Punctured'' Euclidean Black Holes, in which the thermal circle shrinks and becomes a long thin tube, resembling  a wormhole, as in Figure 1. The metric $\tau-\tau$ component and the first derivatives of its logarithm vanish as one approaches the puncture. The asymptotic is shared with the usual Euclidean black hole.
\end{itemize}

In all of these geometries, at least one winding mode of the fundamental string becomes light, and in the last two types of geometries - an entire tower of them becomes light. In string compactifications on some small compact manifold $X$ times $S^1 _{\beta}$, one also encounters effective strings from branes that wrap it, which can be light.

To calculate explicit expressions for the entropy, we will make some assumptions about the regularity of the solutions. These are listed in the subsections below.

 As we saw in the previous section, the entropy of the string and brane modes that comes from explicit $\beta$-dependence is given by
 \begin{equation}
 \label{GeneralEntropy}
 	S =
 	\beta \int d^d x \sqrt{G_D} e^{-2\Phi_D} \beta  \frac{\delta L}{\delta \beta}.
 \end{equation}
 The physical variable that appears in the action of all the extended objects that wind around (or have momentum along) the Euclidean time circle is the proper radius
 \begin{equation}
 	R=\frac{\beta e^{\sigma}}{2\pi}.
 \end{equation}
  This variable appears in the HP action Eq. (\ref{HPAction}). In Eq. (\ref{GeneralDeltaL}), we discussed a general term in the action,
 \begin{eqnarray}
 \label{GeneralDeltaL2}
 \Delta L &=& \left(G_{\tau\tau }\frac{w^2\beta^2}{(2\pi \alpha')^2} \chi_w \chi^* _w\right) ^{n_1}\left(G^{\mu \nu}\partial_{\mu} \chi_w \partial_{\nu} \chi^*_w\right)^{n_2} (\chi_w \chi^*_w)^{n_3} f(\phi_i,\partial_{\mu} \phi_i) \cr &=& \left(\frac{ \beta^2 e^{2\sigma}}{(2\pi )^2} \frac{1}{\alpha'^2} w^2 \chi_w \chi^*_w\right) ^{n_1}\left(G^{\mu \nu}\partial_{\mu} \chi_w \partial_{\nu} \chi^*_w\right)^{n_2} (\chi_w \chi^*_w)^{n_3} f(\phi_i,\partial_{\mu} \phi_i).
 \end{eqnarray}
Thus, the EFT action organizes itself in powers of the proper length $R$ for either a string with compact momentum  or a winding fundamental string. This conclusion generalizes for momentum and winding modes of branes.

The dependence on this variable allows us to make the replacement  $\beta \partial_{\beta}  \to\partial_{\sigma}$ if fluxes are not present~\footnote{In Section 4 of \cite{CMW}, such a replacement was mentioned in the context of computing the energy of a general string tree-level solution.}.
  Therefore, Eq. (\ref{GeneralEntropy}) becomes
 \begin{equation}
 \label{derivative}
 	S =\beta\int_M d^d x ~\sqrt{G_d} e^{-2\Phi_d} \frac{\delta }{\delta \sigma}L.
 \end{equation}
In more generality one should subtract flux terms from Eq. (\ref{derivative}) with non-trivial dependence on $\sigma$, a few examples of such terms will be encountered in the next section.
 The Euler-Lagrange equation for $\sigma$, derived from the EFT then reads:
 \begin{eqnarray}
 \label{EL}
 S&=&  \beta \int_M  d^d x ~ \frac{\delta }{\delta \sigma}\left(\sqrt{G_d} e^{-2\Phi_d} L\right)\nonumber\\
  &=&\beta \int_{\partial M} d^{d-1} x~n_{\mu} \left[ \sqrt{G_d}e^{-2\Phi_d}    \frac{\delta  L}{\delta (\partial_{\mu} \sigma)}-\partial_{\nu} \left(\sqrt{G_d}e^{-2\Phi_d}\frac{\delta L}{\delta (\partial_{\mu} \partial_{\nu} \sigma)}\right)+...\right],\nonumber\\ &
 \end{eqnarray}
 where the term with $n$ derivatives of $\sigma$ is taken when fixing the rest of the derivatives of $\sigma$.

 \subsection{Generalized Horowitz-Polchinski Geometries}

The geometries of interest are asymptotically $S^1 _{\beta}\times R^d$ where $\beta>\beta_H$ and are paramterized by their Einstein-frame ADM mass $M_{E}$. They are horizonless and smooth. Their line element in asymptotic infinity in the Einstein frame, for $D>3$, takes the form
\begin{equation}\label{GHP}
	ds^2 = \left( 1-\frac{2\kappa^2 M_{E}}{\omega_{D-2} (D-2)r^{D-3}}\right)d\tau^2 + \frac{dr^2}{ 1-\frac{2\kappa^2 M_{E}}{\omega_{D-2} (D-2)r^{D-3}}} + r^2 d\Omega_{D-2} ^2 ~,~ r\to \infty.
\end{equation}
The asymptotic form of the dilaton depends on a constant $C_{\phi}$:
\begin{equation}\label{Dilaton}
	\Phi_D = \Phi_0-\frac{C_{\phi}}{r^{D-3}}.
\end{equation}
In the string frame, one has $e^{2\sigma} = e^{\frac{-4\Phi_D}{D-2}} G_{\tau \tau} ^{~\text{Einstein}}$.

We start by computing the entropy coming from the asymptotic part of the boundary $\partial M_{\infty}$.
 Recall that the Lagrangian density of the leading-order dilaton-gravity action contains the term \cite{PolchinskiVolI},
 \begin{equation}
L_{\sigma} = \frac{1}{2\kappa_0 ^2}G _d^{\mu \nu} \partial_{\mu} \sigma \partial_{\nu} \sigma. 	
 \end{equation}
Substituting this term into Eq.~(\ref{EL}),   results in
 \begin{equation}\label{BT}
 	S =\frac{\beta}{\kappa_0 ^2}\int_{\partial M_{\infty}} d^{d-1} x~ e^{-2\Phi_d} \sqrt{G_d} n_{\mu}~ G_d ^{\mu \nu} \partial_{\nu} \sigma.
 \end{equation}
Using Eqs. (\ref{GHP}) and (\ref{Dilaton}),
\begin{equation}\label{GenHPEntropy}
	S=\tfrac{D-3}{D-2}~\beta \left( M_{E}-\frac{2\omega_{D-2}}{\kappa^2} C_{\phi}\right).
\end{equation}
Neither $\alpha'$ correction terms in the action nor terms depending on the matter fields contribute in the asymptotically flat part of the manifold to Eq.~(\ref{EL}) - they vanish because of the vanishing of matter fields at infinity and the asymptotically flat space.

Next, we turn to consider the contribution to the entropy from the surface at $r=0$. We assume that the quantities $e^{-2\Phi_d}n_{\mu}\frac{\delta L}{\delta \left(\partial_{\mu} \sigma\right)}$,~$ n_{\mu}\partial_{\nu} \left(e^{-2\Phi_d} \frac{\delta L}{\delta \left(\partial_{\mu} \partial_{\nu} \sigma\right)}\right)$,...,  are either finite, vanish or diverge slower than $\frac{1}{r^{D-2}}$. A putative stronger divergence would indicate a naked singularity, which is believed to be forbidden.  It follows that the contribution from the origin vanishes and the contribution from the asymptotic boundary $\partial M_{\infty}$ in Eq. (\ref{GenHPEntropy}) captures the entire entropy.

The resulting entropy in Eq. (\ref{GenHPEntropy})  agrees with the entropy derived from the Gibbons-Hawking boundary term \cite{GH}, \cite{Tseytlin88},\cite{KazakovTseytlin},\cite{CMW}, which is valid to all orders in the $\alpha'$ and to leading order in the string coupling. To briefly remind it, using the dilaton equation of motion and adding a GHY term at infinity, the on-shell action $I_{cl}$ is
\begin{equation}
	I_{cl} = -\frac{1}{\kappa_0^2}\int_{\partial M_{\infty}} d^{D-1} x ~n^{\mu} ~\partial_{\mu} \left( \sqrt{h} e^{-2\Phi_D} \right).
\end{equation}
The determinant of the induced metric on the boundary is denoted by $h$.
This can be regularized in the asymptotic part of the manifold. One typically assumes that no contribution arises from an inner boundary of the manifold because, for example, the volume vanishes there while  $n_{\mu}\frac{\delta L}{\delta (\partial_{\mu} \Phi_D)}$ is finite. One can then derive the energy $M = \partial_{\beta} I$ and compare it with the ADM mass, ultimately yielding
\begin{equation}\label{GH}
S= \tfrac{D-3}{D-2} ~\beta\left( M_{E}-\frac{2\omega_{D-2}}{\kappa^2} C_{\phi}\right),
\end{equation}
 for the entropy of the classical solution, to all order in $\alpha'$. It was further suggested in \cite{CMW} that this result is exact in $\alpha'$.

One can view the combination in parenthesis of Eq. (\ref{GenHPEntropy}) or (\ref{GH}) as the ADM mass in the string frame. Technically, $ds^2 _{E}$ and $M_{E}$ in Eq. (\ref{GHP}) are replaced by $ds^2 _{string}$ and the ``string frame ADM mass'' $M_{str}$, respectively. Then, we obtain a version of the Schwarzschild black hole entropy-mass relation $S=\tfrac{D-3}{D-2}\beta M_{str}$ for all the generalized HP solutions.
The numerical factor $C_D^{-1}=\tfrac{D-3}{D-2}$ fixes the normalization of the charge $Q$ so that it is identified with the energy of the strings and branes.

This relation $S=\tfrac{D-3}{D-2}\beta M_{str}$ is interesting in the context of the black hole/string transition \cite{Bowick:1985af},\cite{Susskind93},\cite{HP96} (see also \cite{DamourVeneziano}) - the numerical coefficient in the entropy-mass relation does not vary between the two phases as the energy, defined in the string frame, is changed. This is in contrast to what happens when describing the system using the Einstein's frame, where the Hagedorn entropy of strings transforms into the black hole entropy.

\subsection{Euclidean Black Holes}

We would like to compute contributions to the entropy $S$ of string and brane modes for a string theoretic neutral black hole background whose asymptotic is that of the $\alpha'$-corrected Euclidean Schwarzschild solution, and the near horizon geometry is that of a smooth tip. The goal of this section is to compare these contributions to the Wald entropy and to the Gibbons-Hawking entropy.

The line element near the horizon and at asymptotic infinity takes the form
\begin{equation}\label{LineElement}
	ds^2 = e^{2\sigma (r)} d\tau^2 + e^{-2\sigma(r)} dr^2 + r^2 d\Omega_{D-2} ^2.
\end{equation}
Appendix A explains that there is a coordinate system for which this line-element is valid near the horizon.
On the smooth tip hypersurface $r=r_0$,
\begin{equation}\label{Smooth}
e^{\sigma(r_0)}=0 ~,~	\beta = \frac{2\pi  }{e^{2\sigma}\sigma' (r_0)}.
\end{equation}
We wish to stress that many of the relations derived below are valid for more general manifolds containing an ``interior boundary'' which in the above example is at $r=r_0$.

As in the previous subsection, the asymptotic boundary contribution gives
\begin{equation}\label{BT2}
	S_{asym} =\frac{\beta}{\kappa_0^2}\int_{\partial M_{\infty}}\!\!\!\!\!\!\! d^{d-1} x~ e^{-2\Phi_d} \sqrt{G_d} n_{\mu}~ G_d ^{\mu \nu} \partial_{\nu} \sigma =
		\tfrac{D-3}{D-2}\beta \left( M_{E}-\frac{2\omega_{D-2}}{\kappa^2} C_{\phi}\right)\!. \ \ \ \ \
\end{equation}
$\alpha'$ correction terms as well as matter terms do not modify this.

Next, we would like to compute  contributions to the entropy surface term from the tip. One class of terms originates from the dependence of the action on the Riemann tensor and its covariant derivatives, while another comes from string and brane interaction terms, such as
\begin{eqnarray}
\label{dL2}
	&\Delta L =c_{n_1,n_2,n_3} \kappa_0^{2(n_2+n_3-1)}(\alpha')^{n_1+n_2-1}\left(G^{\mu \nu}\partial_{\mu} \sigma \partial_{\nu} \sigma\right)^{n_1} \times \nonumber\\
	 &\left(G^{\mu \nu}\partial_{\mu} \chi_w \partial_{\nu} \chi_w ^* \right)^{n_2}  \left(\chi_w \chi_w ^*\right)^{n_3}.
\end{eqnarray}
Dimensional analysis determines the powers of $\kappa_0$ and of $\alpha'$. The interaction terms have an overall scaling $\sim\frac{1}{\kappa_0 ^2}$ due to the fields being classical, $\chi_w \propto \frac{1}{\kappa_0}$. The real dimensionless coefficients $c_{n_1,n_2,n_3}$ are formally determined by a string S-matrix calculation. One can also add terms that mix different winding numbers while still preserving the $U(1)$ gauge redundancy.
The contribution to the entropy from the term in Eq.~(\ref{dL2}) is equal to
\begin{eqnarray}
\label{matter}
	\Delta S &=&\frac{\beta}{\kappa_0^2} \ \int_{\partial M} d^{d-1}x  \sqrt{G_d} e^{-2\Phi_d} ~n^{\mu} \partial_{\mu} \sigma~ \times   ~2n_1~ c_{n_1,n_2,n_3} \kappa_0 ^{2(n_2+n_3)}(\alpha')^{n_1+n_2-1} \times   \nonumber \\
	&&(G^{\mu\nu} \partial_{\mu} \sigma \partial_{\nu} \sigma)^{n_1-1}\left(G^{\mu \nu}\partial_{\mu} \chi_w \partial_{\nu} \chi_w ^*\right)^{n_2}  \left(\chi_w \chi_w ^*\right)^{n_3}.
\end{eqnarray}
Unless the matter fields vanish at the tip, we do not expect that the sum of such terms vanishes. We will consider the former possibility as a special case below.

Next, consider the dependence of the action on the Riemann tensor. Using the chain rule, $S$ in Eq. (\ref{EL}) includes terms of the form
 \begin{eqnarray}
 \label{preWald}
 S_R\!\!&=&\!\!\!\! \beta\int_{\partial M}\!\!\!\!\!  d^{d-1} x  \sqrt{G_D} e^{-2\Phi_D}~ n_{\mu} \frac{\delta L}{\delta R_{\alpha \beta \gamma \delta}}\left[ \frac{\delta R_{\alpha \beta \gamma \delta}}{\delta\left(\partial_{\mu} \sigma\right)}-\partial_{\nu}\left( \frac{\delta R_{\alpha \beta \gamma \delta}}{\delta \left(\partial_{\mu} \partial_{\nu} \sigma\right)}\right)\right]\cr &+&
 \beta \int_{\partial M} d^{d-1} x~  n_{\mu} \partial_{\nu}\left(-\sqrt{G_D} e^{-2\Phi_D} \frac{\delta L}{\delta R_{\alpha \beta \gamma \delta}} \right) \frac{\delta R_{\alpha \beta \gamma \delta}}{\delta \left(\partial_{\mu} \partial_{\nu} \sigma\right)}.
 \end{eqnarray}
We show below that this reduces to (minus) the Wald entropy in case the Lagrangian density is independent of the covariant derivatives of Riemann tensor, under assumptions about finiteness that are specified in what follows.

In order to calculate the R.H.S of  Eq. (\ref{preWald}), one must treat the metric component $G_{rr}$ as an independent variable, $G_{rr}= e^{-2\nu(r)}$, and set $\nu(r)=\sigma(r)$ at the end of the calculation.
For the line-element in Eq. (\ref{LineElement}), the non-vanishing components of the first term on the R.H.S. of Eq.~(\ref{preWald})) with $\mu,\nu = r$  are
\begin{equation}\label{Riemann1}
	\frac{\delta R_{r\tau r\tau}}{\delta \sigma'}-\frac{d}{dr}\frac{\delta R_{r\tau r\tau}}{\delta \sigma''} = -e^{2\sigma}\sigma'.
\end{equation}
There are of course three additional permutations, as $R_{r\tau r\tau}=-R_{\tau \tau rr}=R_{\tau r \tau r}=-R_{rr\tau \tau}$. Also:
\begin{equation}\label{Riemann2}
	\frac{\delta R_{\theta\tau \theta\tau}}{\delta \sigma'}-\frac{d}{dr} \frac{\delta R_{\theta\tau \theta\tau}}{\delta \sigma''}= -e^{4\sigma} r.
\end{equation}
Spherical symmetry gives rise to similar terms for  $\frac{\delta R_{\tau \theta_i \tau \theta_i}}{\delta \sigma'}-\frac{d}{dr}\frac{\delta R_{\tau \theta_i \tau \theta_i}}{\delta \sigma''}$ where $\theta_i$ is an angle of the $(D-2)$-sphere.
We now make the following assumptions: 1) The quantity $\sqrt{G_D} e^{-2\Phi_D}\frac{\delta L}{\delta R_{\tau\theta_i \tau\theta_i}}$ is finite, vanishes or diverges slower than $e^{-4\sigma}$ as the tip is approached. Otherwise, the geometry can contain a curvature singularity. Then the vanishing of $e^{4\sigma}$, implies that all of these terms vanish as well. 2) We assume that the derivatives of $e^{-2\Phi_D} \sqrt{G_D}\frac{\delta L}{\delta R_{\alpha \beta \gamma \delta}}$ are finite, vanish or diverge slower than $e^{-2\sigma}$ as the tip is approached. The four permutations of the indices $\{r,\tau,r,\tau\}$ are the only ones for which  $\frac{\delta R_{\alpha \beta \gamma \delta}}{\delta \sigma''}$ are $\mp e^{2\sigma}$, others vanish. Then the last term in Eq. (\ref{preWald}) vanishes at the tip.
Using Eq. (\ref{Smooth}) and taking into account the four permutations of the indices, implies that the only nonvanishing contribution of Eq. (\ref{preWald}), is equal to
 \begin{equation}
 	S_R=   -8\pi \int_{\partial M}   d^{D-2} x~ \sqrt{G_{D-2}}~n_r ~e^{-2\Phi_D} \frac{\delta L}{\delta R_{r\tau r\tau }}.
 \end{equation}
Since the normal points towards the tip (as in any one-dimensional integral) we obtain
\begin{equation}\label{postWald}
	S_R= -S_{Wald}.
\end{equation}
A check that the sign is correct is that the Bekenstein-Hawking entropy comes with a minus sign when evaluating the middle integral of Eq. (\ref{BT2}) at a smooth tip, as in Eq. (\ref{postWald}). So far, the  contributions which we calculated add to:
\begin{equation}\label{Result}
	S= \tfrac{D-3}{D-2}~ \beta \left( M_{E}-\frac{2\omega_{D-2}}{\kappa^2} C_{\phi}\right)-S_{Wald} +\sum \Delta S.
\end{equation}
The notation $\sum \Delta S$ stands for the sum of terms of the form that appeared in Eq. (\ref{matter}).
We therefore conclude that in the absence of string or brane matter, $\sum \Delta S=0$ and $S=0$, then the entropy derived from the Gibbons-Hawking boundary term  and the Wald entropy are equal.

So far, we have only discussed terms that contain powers of the Riemann tensor in the action. The Wald entropy is formally valid also for more general class of terms which contain covariant derivatives of the Riemann tensor. It would be interesting to extend the calculation above to include this class.\footnote{However, we expect that EFTs derived from string theory do not include such terms, as EFTs that do include such terms exhibit hyperbolicity violations on static black hole backgrounds (see, for example,  \cite{Papallo:2017qvl},\cite{Brustein:2017iet}).}

  \subsection{Punctured Euclidean Black Holes}

Here we wish to point out that under some similar assumptions to the ones made in the previous subsection, the inner boundary contribution to the entropy $S_{in}$, vanishes at a puncture. By definition, in this region,
  \begin{equation}\label{Puncture}
  	e^{ \sigma} \to 0  ~,~ \sigma'\to0.
  \end{equation}
First, we assume that at this hypersurface,
$\frac{\delta L}{\delta R_{\alpha \beta  \gamma \delta}}$ and	$\partial_{\nu} \frac{\delta L}{\delta R_{\alpha \beta  \gamma \delta}}$ are both finite. This assumption is likely to hold as its violation would indicate a curvature singularity in a place where the $\tau-\tau$ metric component and the first derivatives of it vanish - see Eq. (\ref{Puncture}).
If indeed both expressions are finite, then the R.H.S of Eqs.~(\ref{Riemann1}) and (\ref{Riemann2}) both  vanish and therefore also the R.H.S of Eq.~(\ref{preWald}). Matter terms as in Eq. (\ref{matter}) similarly vanish if
$
	\left(G^{\mu \nu}\partial_{\mu} \chi_w \partial_{\nu} \chi_w ^*\right)^{n_2}(\chi_w \chi_w ^*) ^{n_3}
$ is finite.
Under these assumptions, we obtain
\begin{equation}
	S_{in} = 0.
\end{equation}
Then the entropy computed from the Gibbons-Hawking boundary term is the exact answer for the contribution from the asymptotic boundary as in Eq. (\ref{BT2}). We emphasize that this is the entropy of strings and branes.

For an asymptotic region of a Schwarzschild black hole with $C_{\phi}=0$ which is connected to the puncture in the interior of the manifold,  the entropy of strings and branes becomes
\begin{equation}
	S=\tfrac{D-3}{D-2}~ \beta M_E =\frac{\omega_{D-2} r_0 ^{D-2}}{4G_N},
\end{equation}
where $r_0$ is the horizon radius of a standard black hole with the same asymptotic as the punctured one.
If the asymptotic region is shared with an $\alpha'$-corrected black hole and it is connected to a puncture in the manifold, we draw a stronger conclusion - that the entire Wald entropy of the $\alpha'$-corrected black hole is of the strings and branes in the punctured geometry.

\section{Entropy of Strings and Branes - Charged Cases}

The goal of this section is to check whether the results of Section 3, which shows that for neutral configurations, the entropy of the black hole is equal to the entropy of strings and branes, are valid also for configurations which do carry charges associated with fluxes.  This is not an automatic extension because, as we will see, the entropy boundary term at infinity which we computed in section 3 does not suffice to reproduce the entire black hole entropy. We are indeed able to show that such an extension is possible for a large class of configurations similar to the ones found by Horowitz and Strominger \cite{HS}. Furthermore, this conclusion holds also when the asymptotic space includes an AdS factor.  Also, Section 2 implies that we compute the associated $U(1)$ charges for the configurations producing fluxes. 

We extend the results of the previous section to asymptotically flat charged black p-branes solutions of Horowitz and Strominger (HS) \cite{HS} which are summarized in Table 1. For these solutions, the entropy was computed by following a Gibbons-Hawking procedure in \cite{Cai}. Here, we compute it in a different and simpler way, which does not require any regularization. The same calculation is applied to the entropy of strings and branes in punctured HS black p-branes.
We assume that back reaction of the sources does not alter the kinetic term of the RR and NS-NS potential for electrically charged NS-NS solutions as well as the ones with an RR flux \footnote{For cases for which there is a radial component to the flux, such as  the self-dual RR flux, the EOM plus a specific gauge choice allow one to express the flux kinetic term as a boundary term at infinity. In such cases, the assumption is not needed.}. The resulting entropy is equal to the Bekenstein-Hawking entropy to leading order in $\alpha'$ of the usual HS solutions. Two additional cases are analyzed as well.
The list of examples which is discussed in this section is:
\begin{itemize}
\item
The 13  HS solutions \cite{HS} with either NS-NS flux or RR flux,
\item
The Horne-Horowitz-Stief solution \cite{HHS} that carries winding charge,
\item
Asymptotically D-dimensional AdS black hole times a D-sphere accompanied by an RR flux (for $D=5$).
\end{itemize}

\subsection{Horowitz-Strominger Black p-Branes}

We start by a brief review of the HS solutions so that the discussion is self-contained.
The HS  solutions are parametrized by the rational numbers $\gamma_r, \gamma_x, \gamma_{\phi}$ that are defined below. Their Euclidean line element is
\begin{eqnarray}
	&ds^2 = \left(1-\left(\frac{r_+}{r}\right)^{D-3}\right)\left[ 1- \left(\frac{r_-}{r}\right)^{D-3}\right]^{\gamma_x-1}d\tau^2 + \frac{\left(1- \left(\frac{r_-}{r}\right)^{D-3}\right)^{\gamma_r}}{1-\left(\frac{r_+}{r}\right)^{D-3}}dr^2+\nonumber \\
	&+ r^2 \left[1- \left(\frac{r_-}{r}\right)^{D-3}\right]^{\gamma_r +1}d\Omega_{D-2} ^2 + \left[1- \left(\frac{r_-}{r}\right)^{D-3}\right]^{\gamma_x} \sum_{i=1} ^p dx_i dx_i.
\end{eqnarray}
The geometries possess an inner horizon at $r=r_-$ and an outer horizon at $r=r_+$ (in Lorentzian signature).
The profile of the dilaton is determined from
\begin{equation}
	e^{-2\Phi} =\left[1-\left(\frac{r_-}{r}\right)^{D-3}\right] ^{\gamma_{\phi}},
\end{equation}
and in addition a $(D-2)$-form flux threads the $S^{D-2}$ part of the geometry:
\begin{equation}
	F = Q \epsilon_{D-2}.
\end{equation}
The fully antisymmetric Levi-Civita symbol $\epsilon_{D-2}$ includes a factor of $\frac{1}{\sqrt{G_{D-2}}}$. An exception occurs for $D=7$ where the flux is self-dual in Lorentzian signature.
A useful parameter is $\alpha$, which by definition appears in the kinetic term of the $(D-3)$-form
\begin{equation}
	I_{\text{flux}} = \int d^{10} x \sqrt{G} \frac{2}{(D-2)!} e^{2\alpha \phi} F^2.
\end{equation}
Then,
\begin{equation}
	\delta \equiv \frac{1}{2\alpha ^2 + (7-D)\alpha+2},
\end{equation}
\begin{equation}
	\gamma_r \equiv \delta (\alpha - 1) -\frac{D-5}{D-3},
\end{equation}
\begin{equation}
	\gamma_x \equiv \delta(\alpha + 1),
\end{equation}
\begin{equation}
	\gamma_{\phi} \equiv -\delta \left(4\alpha + 7-D\right).
\end{equation}
The charge of each solution was found to be
\begin{equation}\label{Charge}
	Q = (D-3)\sqrt{\frac{\delta}{2} (r_+ r_-)^{D-3}}.
\end{equation}
All the 13 HS  solutions are listed in Table 1.
\begin{table}[h]
	\begin{center}
		
		\begin{tabular}{ |l |l |l|l|l| }
			\hline
			ST &  Flux & $M_{\tau r}\times S^{D-2}\times T^p$ & M/E  & $(\alpha,\gamma_r,\gamma_x,\gamma_{\phi})$ \\
			\hline
			All & $H_3$ &  $D=5~,~p=5$ & M & (-1,-1,0,1)\\
			\hline
			Heterotic & $G_2$ & $D=4~,~ p=6$ & M & (-1,-1,0,1)\\
			\hline
			Type IIA  & $F_2$ & $D=4~,~ p=6$ & M & (0,$\frac{1}{2}$,$\frac{1}{2}$,$-\frac{3}{2}$)\\
			\hline
			Type IIA  &$F_4$ & $D=6 ~,~ p=4$  & M & $(0,-\frac{5}{6},\frac{1}{2},-\frac{1}{2})$\\
			\hline
			All & $(* H)_7$& $D=9$, $p=1$&  E & $(1,-\frac{2}{3},1,-1)$ \\
			\hline
			Heterotic &   $(*G)_8$ & $D=10$, $p=0$ &E & $(1,-\frac{5}{7},2,-1)$\\
			\hline
			Type IIA  &  $(*F)_8$ & $D=10~,~p=0$ & E & $(0,-\frac{17}{14},\frac{1}{2},\frac{3}{2})$\\
			\hline
			Type IIA & $(*F)_6$ & $D=8~,~p=2$ & E & $(0,-\frac{11}{10},\frac{1}{2},\frac{1}{2})$\\
			\hline
			Type IIB & $F_5=*F_5$ & $D=7 ~,~ p=3$ & Self-Dual & $(0,-1,\frac{1}{2},0)$. \\
			\hline
		\end{tabular}
		\caption{The HS solutions are listed. RR fields are denoted by $F_p$, the Heterotic two-form is denoted by $G_2$ and the NS-NS three-form is denoted by $H_3$. Solutions that carry a magnetic charge are labeled by ``M'' and electric ones are labeled by ``E''.}
	\end{center}
\end{table}

The inverse temperatures of the solutions are related to the metric functions \cite{GH},\cite{VisserDirtyBH},\cite{Cai}:
\begin{equation}
\label{beta}
	\beta = \frac{4\pi \sqrt{G_{\tau \tau} G_{rr}}}{G_{\tau \tau} ' (r_+)}=\frac{4\pi r_+}{D-3} \left(1-\left(\frac{r_-}{r_+}\right)^{D-3}\right)^{\frac{1+\gamma_r - \gamma_x}{2}}.
\end{equation}
For most of the solutions, the ``string frame area'' of the outer horizon is
\begin{equation}
\label{GeneralizedArea}
A_H=	\int_{hor} \sqrt{G_{D-2}} e^{-2\Phi_D}= \omega_{D-2} \text{Vol}(T^p) r_+ ^{D-2}\left(1-\left(\frac{r_-}{r_+}\right)^{D-3}\right)^{\frac{D-1}{2(D-3)}}.
\end{equation}
Exceptions occur for $\alpha=-1,D=4$, for which the rightmost factor  is $\left(1-\left(\frac{r_-}{r_+}\right)^{D-3}\right)^{1}$ and the $\alpha=1,D=10$ case which has $\left(1-\left(\frac{r_-}{r_+}\right)^{D-3}\right)^{\frac{1}{7}}$ instead.

Now we consider replacing the near outer horizon region of the HS solutions, by a long thin tube with a puncture, without changing asymptotic infinity. In particular, the asymptotic circumference $\beta$ at infinity is as in the Euclidean version of the HS solutions.

The boundary term in Eq. (\ref{BT}) associated with the HS backgrounds is given by:
\begin{equation}\label{BT123}
	S_{BT} = \frac{(D-3)\beta \text{Vol}(T^p) \omega_{D-2} r_+ ^{D-3}}{2\kappa^2}\left(1+(\gamma_x-1)\left(\frac{r_-}{r_+}\right)^{D-3}\right).
\end{equation}
Substituting Eq.~(\ref{beta}) into Eq.~(\ref{BT123}),
\begin{equation}
	S_{BT} = \frac{ \text{Vol}(T^p) \omega_{D-2} r_+ ^{D-2}}{4G_N}\left(1+(\gamma_x-1)\left(\frac{r_-}{r_+}\right)^{D-3}\right)\left(1-\left(\frac{r_-}{r_+}\right)^{D-3}\right)^{\frac{1+\gamma_r - \gamma_x}{2}}.
\end{equation}
In the cases of the $H_3\neq 0$ flux in the type II superstring, the Heterotic string and the gauge field flux $G_2\neq 0$ in the Heterotic string (these appear in the first two lines of table 1), this is the only contribution to leading order in $\alpha'$ and the string coupling. Setting $\gamma_r=-1~,~ \gamma_x =0$, and comparing with Eq. (\ref{GeneralizedArea}), the conclusion is that the entropy of the string and brane modes is given by
\begin{equation}
\label{salpha-1}
	S (\alpha=-1) = \frac{{A_H}}{4G_N}.
\end{equation}
Equation~(\ref{salpha-1}) agrees with the results in \cite{Cai} which were calculated in the Einstein frame. In addition, at extremality one can obtain zero both from the calculation above, and by computing the same boundary term for another slicing of the geometry where the $\tau-\tau$ component of the metric is constant \cite{HS}.

Next, consider RR fluxes in type IIA or IIB, in which case the above
\begin{equation}\label{RRrelations}
	\alpha=0~,~ \delta=\frac{1}{2}~,~ \gamma_x = \frac{1}{2} ~,~ \gamma_r = -\frac{3D-13}{2(D-3)}.
\end{equation}
The relevant $\sigma$ EOM, with a non-standard normalization of the RR kinetic term, is the following
\begin{equation}\label{RREq}
	\frac{1}{\kappa^2}\partial_{\mu} \left(e^{-2\Phi_D} \sqrt{G_D} \partial^{\mu} \sigma\right) = \sqrt{G_d} \left[\frac{e^{\sigma}}{\kappa^2 (D-2)!} F_{D-2}^2 + s_d\right].
\end{equation}
To explain the normalization, the action contains the two terms $-\frac{e^{-2\Phi_D}R_D}{2\kappa^2}+\frac{|F_{D-2}|^2}{\kappa^2(D-2)!}$.~\footnote{Note that a factor $\frac{1}{4\kappa^2} |F_p|^2$ appears in Polchinski II Eqs. (12.1.10c), (12.1.26c).}  String and brane sources are included in $s_d$ and Eq. (\ref{EL}) ties them to both the dilaton-gravity boundary term and the flux term in Eq. (\ref{RREq}).

The ``flux term'' which contributes to the entropy is
\begin{equation}\label{FT}
	S_{FT} = -\frac{\text{Vol}\left(T^p \right) \omega_{D-2}\beta}{\kappa^2 }\int \sqrt{G_D} \frac{1}{(D-2)!}F_{D-2} ^2 dr.
\end{equation}
This can be evaluated by plugging the HS solution,
\begin{equation}
	S_{FT} =\frac{(D-3)\text{Vol}(T^p) \omega_{D-2} \beta r_+ ^{D-3}}{2\kappa^2 }\left(-\frac{r_- ^{D-3}}{2r_+ ^{D-3}}\right).
\end{equation}
For $D=7$ one can check that the resulting flux term is identical.
Using the value of  $\beta$ in Eq.~(\ref{beta}), the resulting  entropy
\begin{equation}
	S(\alpha=0)=S_{BT}+S_{FT} = \frac{{A_H}}{4G_N},
\end{equation}
is exactly equal to the Bekenstein-Hawking entropy - as can be seen from Eq.~(\ref{GeneralizedArea}). This result also agrees with \cite{Cai}.

Next, consider the $\alpha=1$ $(*H)_7$ and $(*G)_8$ electric black string and black hole. The values of the parameters are:
\begin{equation}
	p=10-D ~,~ \delta = \frac{1}{11-D} ~,~ \gamma_x = \frac{2}{11-D} ~,~ \gamma_r = -\frac{D-5}{D-3} ~,~ \gamma_{\phi} = -1.
\end{equation}
The relevant EOM is the following,
\begin{equation}
	\frac{1}{\kappa^2}\partial_{\mu} \left(e^{-2\Phi_D} \sqrt{G_D} \partial^{\mu} \sigma\right) = \sqrt{G_d} \left[2\frac{e^{2\sigma+2\Phi_d}}{\kappa^2 (D-2)!} (*H)^2 _{D-2} + s_d\right].
\end{equation}
The flux term is equal to
\begin{eqnarray}
	S_{FT}=
	-\frac{ (D-3)\beta \omega_{D-2} \text{Vol}(T^p)r_+ ^{D-3} r_- ^{D-3}}{2\kappa^2}\frac{2}{(11-D)}\left(\frac{1}{r_+}\right) ^{D-3}.
\end{eqnarray}
The parameter $D$ takes the values $9,10$. It follows that
\begin{equation}
	S=S_{BT}+S_{FT} = \frac{\text{Vol}(T^p)\omega_{D-2} r_+ ^{D-2}}{4G_N}\left(1-\left(\frac{r_-}{r_+}\right)^{D-3}\right)^{\frac{1+\gamma_r - \gamma_x}{2}+1},
\end{equation}
which for both $D=9,10$, yields exactly the Bekenstein-Hawking entropy\footnote{See Eq. (\ref{GeneralizedArea}) and the comment below it.}
\begin{equation}
	S(\alpha=1)= \frac{A_H}{4G_N}.
\end{equation}
This agrees with \cite{Cai}.

\subsection{Black p-Branes with a Winding Charge}

We consider the Horne, Horowitz and Stief solution \cite{HHS} which includes a black p-brane with a winding $B_2$ charge. The winding is about a spatial dimension $x$ of circumference $2\pi R$, and the winding number is determined by the charge. The line-element, the exponential of minus twice the dilaton and the B-field are given by Eq.~(16) in \cite{HHS}. We would like to translate it to our conventions by continuing to Euclidean signature $\tau=it$, and performing the following replacements
\begin{equation}
	n\to D-3 ~,~M_0 \to r_0 ^{D-3}~,~ \phi \to -2\Phi_D,
\end{equation}
\begin{eqnarray}
	ds^2 &=& \frac{\left( 1-\left( \frac{r_0}{r}\right)^{D-3}\right)}{1+\left(\frac{r_0}{r}\right) ^{D-3} \sinh^2 (\alpha)}d\tau^2 + \frac{dr^2}{1-\left( \frac{r_0}{r}\right)^{D-3}} \cr &+&\frac{dx^2}{1+\frac{r_0 ^{D-3}}{r^{D-3}}\sinh^2 (\alpha)} +r^2 d\Omega_{D-2} + \sum_{i=1} ^p dx_i dx_i,
\end{eqnarray}
\begin{equation}
	e^{-2\Phi_D} =1+\frac{r_0 ^{D-3}}{r^{D-3}} \sinh^2 (\alpha)
\end{equation}
and
\begin{equation}
	B_{x\tau} =-i\frac{r_0 ^{D-3} \cosh(\alpha) \sinh(\alpha)}{r^{D-3}+r_0 ^{D-3} \sinh^2 (\alpha)}.
\end{equation}
We now apply a gauge transformation which renders $B_{x\tau}=0$ at the inner boundary, by adding $+i \tanh(\alpha)$. The three-form flux is simply the radial derivative of the B-field.
The area of the horizon can again be evaluated,
\begin{equation}
A_H=	\int_{hor} \sqrt{G_{d-1}} e^{-2\Phi_D} = \text{Vol}(T^p) \omega_{D-2}  2\pi R r_0 ^{D-2} \cosh(\alpha),
\end{equation}
as can the inverse temperature associated with a smooth tip,
\begin{equation}
	\beta = \frac{4\pi \sqrt{G_{\tau \tau }G_{rr}}}{G_{\tau \tau}'} |_{r=r_0} = \frac{4\pi r_0}{D-3} \cosh(\alpha).
\end{equation}
As before, we consider the punctured version of the solution.
The gravity boundary term for the entropy at infinity gives
\begin{eqnarray}
	S_{BT}&=& \frac{\text{Vol}(T^p) \omega_{D-2} \beta 2\pi R r_0 ^{D-3}}{2\kappa^2}(D-3)\left( 1+\sinh^2 (\alpha)\right) \cr &=& \frac{A_H}{4 G_N} \left( 1+\sinh^2 (\alpha)\right).
\end{eqnarray}
The flux kinetic term involves the numerical factor $\frac{1}{12}$ and includes a summation over 6 equal permutations of the indices $r,x,\tau$.  In the $\sigma$ EOM, this term should be added to $S_{BT}$ with an overall minus sign due to a) the factor of $e^{-2\sigma}$ in the Lagrangian density $e^{-2\Phi_d -2\sigma} G^{rr} G^{xx} H_{r\tau x }^2$ and b) the factor of $i^2$. This term can be represented as a boundary term at infinity:
\begin{eqnarray}
	&S_{FT} = -\frac{\text{Vol}(T^p) \omega_{D-2} \beta ~2\pi R}{2\kappa^2} G^{rr} G^{xx} e^{-2\sigma} \sqrt{G_D}e^{-2\Phi_D} B_{x\tau } \frac{d}{dr} B_{x\tau} |^{\infty} _{r_{0}}=\nonumber\\
	&-\frac{A_H} {4G_N} \sinh^2(\alpha).
\end{eqnarray}
The sum $S_{BT}+S_{FT}=\frac{A_H}{4 G_N} $, exactly the Bekenstein-Hawking entropy. The T-dual solution with momentum charge works in the same way.

\subsection{Black Hole in AdS}

Next, consider a black hole solution in asymptotically AdS$_D \times S^D$ with $D=5$. The AdS length scale is denoted by $R$. The asymptotic line-element reads
\begin{eqnarray}
	&ds^2 = f(r)d\tau ^2 + \frac{dr^2 }{f(r)}+r^2 d\Omega_{D-2} ^2 +R^2 d\Omega_{D} ^2 ~,~ r\to \infty,
\end{eqnarray}
\begin{equation}
	f(r)\equiv \frac{r^2}{R^2}+1-\left(\frac{r_0}{r}\right)^{D-3}\left(\frac{r_0 ^2}{R^2}+1\right) ~,~ r\to \infty.
\end{equation}
We assign the $AdS_D$ the standard asymptotic periodicity:
\begin{equation}\label{BetaAdS}
	\beta = \frac{4\pi r_0}{D-3 +(D-1) \frac{r_0 ^2}{R^2}}.
\end{equation}
The RR flux is imaginary self-dual:
\begin{equation}
	F = f \left(\epsilon_{S^D}+i\epsilon_{AdS_D}\right).
\end{equation}
The factor of $i$ appears because we analytically continued the original Lorentzian solution to Euclidean signature. The metric EOM in the asymptotic AdS region implies that
\begin{equation}\label{EOMmetric}
	\frac{D-1}{R^2}=\frac{1}{8} f^2 e^{2\Phi_D}.
\end{equation}
The $\sigma$ EOM in the presence of possible sources is given by
\begin{equation}
	\frac{1}{\kappa_0 ^2}\partial_{\mu} \left(e^{-2\Phi_D} \sqrt{G_D} \partial^{\mu} \sigma\right) = \sqrt{G_d} \left[-\frac{e^{\sigma}}{8\kappa_0 ^2 D!} F_{AdS_D}^2 + s_d\right].
\end{equation}
We replace the region near the tip by a long thin tube with puncture boundary conditions.
The entropy boundary term Eq. (\ref{BT}) at some cutoff hypersurface $r=r_c$ is
\begin{equation}
	S_{BT} = \frac{\text{Vol}(S^{D}) \omega_{D-2}\beta }{2\kappa_0^2 e^{2\Phi_D}}r _c ^{D-2}\left(\frac{2r_c}{R^2}+\frac{(D-3) r_0 ^{D-3}}{r_c ^{D-2}}\left(\frac{r_0 ^2}{R^2}+1\right)\right).
\end{equation}
The flux term is given by
\begin{equation}
	S_{FT} = -\frac{\text{Vol}\left( S^D\right) \omega_{D-2}\beta}{8\kappa_0 ^2}\int_{r_0} ^{r_c} dr~ r^{D-2} f^2.
\end{equation}
Using Eq. (\ref{EOMmetric}),
\begin{equation}
	S_{FT} = -\frac{\text{Vol}\left( S^D\right) \omega_{D-2}\beta}{\kappa_0^2 e^{2\Phi_D}  R^2} \left(r_c ^{D-1}-r_0 ^{D-1}\right).
\end{equation}
Summing the two terms $S_{BT}+S_{FT}$ and using Eq. (\ref{BetaAdS}), one obtains
\begin{equation}
	S =S_{BT}+S_{FT} = \frac{A_H}{4G_N}.
\end{equation}

\section{Conclusions and Discussion}

In this paper we discussed the thermodynamic entropy of strings and branes that wrap or vibrate along the thermal circle in various geometries.

We found that the Noether charge $Q$, associated with translations along the T-dual of the thermal circle and the entropy of winding strings $S$,  are related by $S\propto \beta Q$. This is a similar relation to the relation between the Noether charge associated with translations along the Killing horizon and the Wald entropy. However, we pointed out that in spite of this similarity, there are significant differences. Importantly, $Q$ is sourced exclusively by wrapped strings and branes and vanishes in their absence.

We demonstrated that the entropy is only sensitive to the behavior of the fields at the boundary of space and consequently, once the asymptotic behavior of the solution is specified, it is less sensitive to corrections to the EFT  action  than one would expect. We also checked, under certain assumptions of regularity, that for neutral black holes solutions of pure higher derivative theories of gravity, the Wald entropy is equal to the entropy derived from Gibbons-Hawking boundary term.  Furthermore, we showed that  for generalized HP solutions and neutral black hole solutions with a puncture, for which the entropy boundary term does not receive contributions from the inner boundary, the entropy, to all orders in $\alpha'$, is accounted for by the entropy of the strings and branes.

We further argued that if charged black holes satisfy the puncture boundary conditions, then  the Bekenstein-Hawking entropy of the standard Horowitz-Strominger solutions is reproduced by the entropy of strings and branes, to leading order in $\alpha'$. One can alternatively view these calculations as shortcuts to performing the Gibbons-Hawking procedure, which do not require a regularization.

One can reverse the logic which we followed in this paper by assuming that the entropy of the strings and branes is equal to the entropy of the black hole to all orders in $\alpha'$, and ask what are the conditions on the inner boundary of the manifold. Two possible answers emerge, either a punctured geometry or  a geometry for which the volume of an $S^{D-2}$ at the origin shrinks to zero, while both asymptote to an $\alpha'$-corrected black hole solution. In both cases, the inner boundary is not a standard horizon and the geometry is non-singular. The absence of a horizon is consistent with the general pattern found in the Fuzzball program, that a horizon results from an insufficient inclusion of stringy effects \cite{Samir},\cite{FuzzballReview}. One can view this class of solutions as corresponding to the state of the black hole when the string sources are included explicitly, with a string scale resolution. When these sources are integrated out, the result is a geometry with a horizon. The entropy of the black hole in both descriptions needs to be evaluated using different methods, which nevertheless lead to the same value of the entropy.

The result for the punctured black holes is related to the FZZ duality \cite{KazakovKostovKutasov}. This duality implies that the entropy of the winding condensate on the cylinder with a potential wall is equal to the entropy of the cigar. Similarly, we argued that the latter is equal to the entropy of strings and branes in another geometry, where the thermal cycle does not pinch off.

There are several ways in which our results can be generalized:
\begin{itemize}
\item
It would be interesting to extend our analysis to other types of inner boundaries. For instance, one can imagine cutting a disk from the cigar at some $r=r_0+\epsilon$ with $\epsilon \ll r_0$ and compute the inner boundary contribution to the entropy of strings and branes as a function of $\epsilon$. A different boundary which would be interesting to study is a tip with a conical singularity, which was recently discussed in \cite{Atish2}. Another example to consider would be the Euclidean version of de-Sitter spacetime.
\item
An interesting question is what fraction of the entropy does each string and brane condensate carry? We expect that for solutions without fluxes, the winding modes of winding number $\pm 1$ carry most of the entropy because they are lighter than the other modes in a larger region of the manifold. For such solutions with a small string coupling throughout the manifold, branes are expected to carry a tiny fraction of the entropy because they are heavy.
\item
Another interesting question is whether a near-puncture region can be embedded in string theory. This could be answered by attempting to construct an appropriate 2D worldsheet superconformal field theory.
\item
Finally, our results could be complemented by mapping our Euclidean, target-space calculations to  Lorentzian, CFT calculations as in \cite{StromingerVafa}. We expect that the entropy is equal to the logarithm of the number of Lorentzian microstates with the same macroscopic energy and charges.  We also expect the total length of a string to be related in a simple way to its entropy in a weakly-coupled string theory.
\end{itemize}

\section*{Acknowledgements}

We would like to thank Yiming Chen, Amit Giveon, Sunny Itzhaki, David Kutasov, Emil Martinec and Samir Mathur for useful discussions and many comments on this work. We especially thank Ofer Aharony for comments and questions. We would also like to thank Kostas Skenderis for communications. The work of RB and YZ is supported by the German Research Foundation through a German-Israeli Project Cooperation (DIP) grant ``Holography and the Swampland.'' YZ  is supported by the Adams fellowship.
%\begin{spacing}{1.25}

\appendix
\section{Justifying the Near-Horizon Metric Choice}
In this appendix we wish to show that there is a coordinate system for which the metric in Eq. (\ref{LineElement}) is valid near the tip.
One starts with
\begin{equation}\label{LineElement0}
	ds^2 = e^{2\sigma (r)} d\tau^2 + e^{-2\nu(r)} dr^2 + r^2 d\Omega_{D-2} ^2.
\end{equation}
Here, $\nu(r)$ is a function which is generically different from $\sigma(r)$. The existence of a smooth tip implies that
\begin{equation}
	e^{\sigma(r_0)}=0 ~,~ \beta = \frac{2\pi e^{\sigma-\nu}}{e^{2\sigma} \sigma'} (r=r_0).
\end{equation}
Taking the near-horizon limit allows one to approximate the line-element by fixing the prefactor of the angular coordinates:
\begin{equation}\label{LineElement2}
	ds^2 = e^{2\sigma (r)} d\tau^2 + e^{-2\nu(r)} dr^2 + r_0^2 d\Omega_{D-2} ^2.
\end{equation}
Next, we can apply a diffeomorphism transformation:
\begin{equation}
	dr=e^{\nu(r)- \sigma(r)}d\tilde{r}.
\end{equation}
This brings about the following metric,
\begin{equation}\label{LineElement3}
	ds^2 = e^{2\sigma (\tilde{r})} d\tau^2 + e^{-2\sigma(\tilde{r})} d\tilde{r}^2 + r_0^2 d\Omega_{D-2} ^2.
\end{equation}
This coordinate system still admits a tip because the original $r=r_0$ is mapped to some $\tilde{r}_0$ at which $e^{\sigma}$ vanishes, this just relabels the position of the tip. Also, the asymptotic circumference $\beta$ remains unchanged. Utilizing the chain rule, one readily checks that in this coordinate system, the tip is still smooth:
\begin{equation}
	\beta = \frac{2\pi e^{\sigma-\nu}}{e^{2\sigma} \frac{d}{dr}\sigma}(r=r_0)=\frac{2\pi }{e^{2\sigma} \frac{d}{d\tilde{r}} \sigma (\tilde{r}_0) }.
\end{equation}
Finally, we replace $\tilde{r}\to r$ in the body of the paper to avoid cluttering. We thus conclude that near the horizon, there is a coordinate system for which Eq. (\ref{LineElement}) is valid.
\linespread{1.15}\selectfont

%\end{spacing}
\end{document}